\documentclass{elsart}
\usepackage[dvips]{epsfig}
\usepackage{amssymb}

\newcommand{\be}{\begin{equation}}
\newcommand{\ee}{\end{equation}}
\newcommand{\bea}{\begin{eqnarray}}
\newcommand{\eea}{\end{eqnarray}}
\newcommand{\Dslash}{\mbox{$D$\kern-0.65em \hbox{/}\hspace*{0.25em}}}
\newcommand{\Dslashup}{\not \kern-0.03em D}
\renewcommand{\det}{{\rm det}}
\newcommand{\Langle}{\raise.16ex\hbox{$\langle$}}
\newcommand{\Rangle}{\raise.16ex\hbox{$\rangle$}}
\newcommand{\lap}{\lesssim}

\begin{document}


\begin{frontmatter}
\title{The MultiBoson method}
\author{Philippe de Forcrand\thanksref{email}}

\address{ETH, CH-8092 Z\"urich, Switzerland}
\thanks[email]{\tt forcrand@scsc.ethz.ch}

\begin{abstract}
This review describes the multiboson algorithm for Monte Carlo
simulations of lattice QCD, including its static and dynamical aspects,
and presents a comparison with Hybrid Monte Carlo. 
\end{abstract}
\end{frontmatter}

\section{The MultiBoson approach in a nutshell}

Monte Carlo simulations of lattice QCD aim at sampling the partition function
\be
{\mathcal Z} = \int \prod_{x,\mu} dU_{x,\mu}~e^{-S_g(\{U\})} ~ 
\prod_{i=1}^{N_f} \det(\Dslash(\{U\}) + m_i) \quad,
\ee
where $U_{x,\mu}$ are gauge links, $S_g(\{U\})$ is the {\em local} gauge
action (e.g. plaquette), $\Dslash(\{U\})$ is the {\em local} discretized
Dirac operator, and the fermion (quark) of flavor $i$ has mass $m_i$,
$i=1,..,N_f$. The determinant results from the integration over the
anti-commuting fermion fields. It couples all gauge links $U$ to each other.
Such a non-local interaction implies that, in order to perform the Monte
Carlo update of a single gauge link $U_{x,\mu} \rightarrow U'_{x,\mu}$,
one must compute its interaction with all other links, which represents an
amount of work at least ${\mathcal O}(V)$, where $V$ is the lattice volume.
Updating all links costs ${\mathcal O}(V^2)$ or more, in contrast to the case
of a local interaction like $S_g$, where the cost is simply ${\mathcal O}(V)$.
Therein lies the difficulty of simulating dynamical fermions.

Progress in this old problem came with the introduction of an
auxiliary boson field $\phi$ and the use of the Gaussian integral 
representation, appropriate for the case of two flavors degenerate in mass:
\be
|\det (\Dslash + m)|^2 = \int [d\phi^\dagger] [d\phi]~
e^{- |(\Dslashup + m)^{-1} \phi|^2} \quad.
\label{1.2}
\ee
This identity is at the core of the Hybrid Monte Carlo (HMC) method
\cite{ADK}. One alternates updates of $\{U\}$ and $\{\phi\}$. However,
the effective action $|(\Dslash + m)^{-1} \phi|^2$ is {\em still} 
non-local because $(\Dslash + m)^{-1}$ is. The idea is that the work
${\mathcal O}(V)$ previously necessary to update one link $U_{x,\mu}$
by an arbitrary step can now be expended to update {\em all} links
$U$ by an infinitesimal (in fact, very small) step (see \cite{ADK} for details).

To get rid of non-locality altogether, L\"uscher proposed \cite{Bible}
to turn Eq.(\ref{1.2}) ``upside-down'':
\be
|\det (\Dslash + m - z_k)|^{-2} = \int [d\phi^\dagger] [d\phi]~
e^{- |(\Dslashup + m - z_k) \phi|^2} \quad,
\label{1.3}
\ee
which is true for any complex number $z_k$ outside the spectrum of 
$(\Dslash + m)$. The action in the right-hand side is now local. Its range
is twice that of $\Dslash$: $\phi_x$ interacts at most with its 
next-nearest neighbors for the standard discretization of $\Dslash$
(Wilson or staggered). On the left-hand side, one can approximate the 
desired $|\det (\Dslash + m)|^2$ by considering the product of factors
Eq.(\ref{1.3}) with different $z_k$'s, each with its own auxiliary boson
field $\phi_k$. If one chooses a polynomial $P_n(x) = c_n \prod_{k=1}^n
(x - z_k)$ which approximates $1/x$ over the whole spectrum of 
$(\Dslash + m)$, then $P_n(\Dslash + m) \approx (\Dslash + m)^{-1}$, and
\bea
\lefteqn{
|\det (\Dslash + m)|^2 \approx |\det^{-1} P_n(\Dslash + m)|^2  
= c_n^V \prod_k |\det^{-1} (\Dslash + m - z_k)|^2 }  \nonumber \\
&~~~~~~~~~~~= c_n^V \int \prod_k [d\phi_k^\dagger] [d\phi_k]~
e^{- \sum_k |(\Dslashup + m - z_k) \phi_k|^2} \quad.
\eea
This equation summarizes the MultiBoson (MB) method, with effective action
\be
S = \sum_{k=1}^n |(\Dslash + m - z_k) \phi_k|^2 \quad. 
\label{MBaction}
\ee
Each term in this action is most sensitive to fluctuations in the Dirac spectrum
nearest $z_k$, so that one may view each boson field $\phi_k$ as 
``dedicated'' to the control of a particular region (IR, UV, ...) of the
spectrum.
Note that the discrete sum over $k$ can be seen as an approximation to
an integral in a fifth dimension. It is interesting that one recovers a very
similar effective action when bosonizing continuum QCD, and {\em then} 
discretizing \cite{Slavnov}.

The MB effective action is local, but it involves $n$ auxiliary boson fields
$\phi_k$. It therefore now takes work ${\mathcal O}(n)$ to update one link $U$
by an arbitrary step. Larger $n$ means more work, but also a more accurate
approximation of $\det (\Dslash + m)|^2$. One clearly wants to choose the
polynomial $P_n$ so as to minimize $n$ while preserving sufficient
accuracy. This is the {\em static} setup of the MultiBoson method, reviewed
in Sec. 2. For the actual MB simulation, one must also arrange the update
of all the gauge and boson fields, so as to explore phase space with minimum 
computer effort. These MB {\em dynamics} issues are reviewed in Sec. 3.
A somewhat subjective assessment of the MultiBoson method and a comparison
with HMC follows in Sec. 4.

In the rest of the paper, the Wilson discretization of the Dirac operator
is normally used, so that $(\Dslash + m)$ is replaced by 
$({\bf 1} - \kappa M)$, where $\kappa$ is the hopping parameter and $M$
the hopping matrix. But the staggered (Kogut--Susskind) Dirac discretization
could be used just as well. In fact, the polynomial approximation which forms
the basic idea of the MB method is quite general. One may construct the
polynomial $P_n(x)$ to approximate an arbitrary function of $x$, like
$1/\sqrt{x}$ or $x^{-1/4}$. This has led to simulations of QCD with 1 flavor
\cite{Nf1} or of $N=1$ supersymmetric theory \cite{SUSY}, respectively. 
The MB method
has even been used in a particularly difficult context, to simulate the
$(2+1)d$ Hubbard model at half-filling \cite{R29}. All these projects
are summarized chronologically in Table I, which tries to sort out the
many avatars of the MB method reviewed next.

\begin{table*}
\caption{Chronological summary of published MultiBoson simulations}
\vspace*{0.3cm}
\label{compare} 
\begin{center}
\begin{tabular}{| c | c | c | c | c | c | c |}
\hline 
Ref. & Theory & \multicolumn{3}{c |}{MultiBoson variant:} & Fermion mass: & Measured\\
and &         & \multicolumn{3}{c |}{} & light (LL) to & improvement \\
\cline{3-5}
Year &        & exact? & non-Herm.? & $\Delta\beta\neq0$? & heavy (HH)& 
over HMC \\
\hline 
\cite{LAT94} 1994 & $SU(2), N_f=2$ &   &   & & LH & --- \\
\hline 
\cite{R78} 1994 & $SU(2), N_f=2$ &   &   & & LH & --- \\
\hline 
\cite{R77} 1994 & $\rm{QED}_2, N_f=2$ & X$^{\ast}$ &   & & LH & --- \\
\hline 
\cite{R75} 1995 & $QCD, N_f=2$ & X$^{\ast}$ &   & & H & --- \\
\hline 
\cite{R29} 1995--97 & $(2+1)d$ Hubbard &   &   & & ``LL''$^{\dagger}$ & --- \\
\hline 
\cite{R66} 1995 & $QCD, N_f=2$ & X & X & & H & ${\mathcal O}(1)$ \\
\hline 
\cite{SUSY} 1995--99 & $N=1$ SUSY & X &   & & LH & --- \\
\hline 
\cite{R60} 1995 & QCD, $N_f=2$ &   &   & & LH & ---  \\
\hline 
\cite{R59} 1995 & QCD, $N_f=2$ &   &   & & H & $\lap 1^{\ddagger}$ \\
\hline 
\cite{R57} 1996 & QCD, $N_f=2$ & X & X & & H & --- \\
\hline 
\cite{R43} 1996 & $\rm{QED}_2, N_f=2$ &   &   & & LH & --- \\
\hline 
\cite{R39} 1996 & QCD, $N_f=2$ & X & X & & L & ${\mathcal O}(1)$\\
\hline 
\cite{Nf1} 1996--98 & QCD, $N_f=1$ & X & X & & HH & ${\mathcal O}(10)$ \\
\hline 
\cite{R34} 1996 & $SU(2), N_f=2$ &   &   & & LH & ${\mathcal O}(1)^{\ddagger}$ \\
\hline 
\cite{R25} 1997 & QCD, $N_f=2$ & X & X & & LH & $\lap 1$ \\
\hline 
\cite{R20} 1997 & $\rm{QED}_2, N_f=2$ & X &   & & LH & 2-3 \\
\hline 
\cite{UV} 1998 & QCD, $N_f=2$ & X & X & X & LH & ${\mathcal O}(5)$ \\
\hline 
\end{tabular}
\vspace*{0.3cm}
\begin{flushleft}
$^\ast$ The correction factor was computed exactly by Lanczos. \\
$^\dagger$ The condition number of the fermion matrix is very large $\sim e^\beta$. \\
$^\ddagger$ Comparison with the Kramers algorithm.
\end{flushleft}
\end{center}
\end{table*}


\section{Statics}
\subsection{How to choose the approximating polynomial?}
\label{nbfields1}
No less than four proposals have been made for selecting the polynomial
$P_n(\Dslash + m)$. The reason is that what constitutes an ``optimal''
polynomial has been a rather subjective issue until the most recent proposal.
The successive proposals have tried to reduce the degree $n$ of the polynomial
for a given ``accuracy.'' The general idea is that $n$ can be reduced 
if one knows more about the spectral properties of the Dirac matrix.
The trade-off is that more adaptive methods do not have the aesthetic
appeal and the clarity of analytic results. We review these proposals
in chronological order, which is also that of increasing accuracy.

\subsubsection{Hermitian Chebyshev polynomial}
\label{herm}

The original proposal of L\"uscher \cite{Bible} 
is to work with the hermitian matrix
\be
Q \equiv \gamma_5 \frac{({\bf 1} - \kappa M)}{(1 + 8\kappa)},
\label{norm}
\ee
where the normalization ensures that the spectrum $\{\lambda\}$ is in $[-1,+1]$.
For $N_f=2$ flavors, the approximating polynomial 
$P_n(Q^2) \approx \frac{1}{Q^2}$
can be chosen as that which minimizes the maximum 
relative error
\be
R(\lambda^2) = \frac{P_n(\lambda^2) - 1/\lambda^2}{1/\lambda^2}
= \lambda^2 P_n(\lambda^2) - 1
\label{eq2}
\ee
in absolute value over the interval $\lambda^2 \in [\varepsilon,1]$.
This polynomial is 
\be
P_n(x) = \frac{1}{x} \left( 1 - 
\frac{T_{n+1}(  \frac{2 x}{1 - \varepsilon}
                     - \frac{1 + \varepsilon}{1 - \varepsilon})}
            {T_{n+1}(- \frac{1 + \varepsilon}{1 - \varepsilon})}
\right),
\ee
where $T_n(x)$ is the $n$th-order Chebyshev polynomial 
[$T_n(cos(\theta)) = cos(n \theta)$].
The great advantage of this choice is that an analytic upper bound on the
error $R(\lambda^2)$ is known:
\be
| R(\lambda^2) | ~\leq 2~
\left(\frac{1 - \sqrt{\varepsilon}}
           {1 + \sqrt{\varepsilon}}\right)^{n+1},
~~\forall~\lambda^2 \in [\varepsilon,1] \quad.
\label{eq1}
\ee
Therefore, it is sufficient to monitor the smallest eigenvalue of $Q^2$.
One can then choose $\varepsilon$ and $n$ so as to {\em guarantee}
any prescribed accuracy.

The main drawbacks of this choice are: $(i)$ the upper bound of $+1$ for
the spectrum of $Q^2$ is very conservative (it is reached in the free 
field only), so that the approximation extends uselessly in a region 
where it is not needed [this can easily be fixed by also monitoring the
largest eigenvalue of $Q^2$ and changing the normalization 
in Eq.(\ref{norm}) accordingly]; 
$(ii)$ it deals with $\gamma_5 \Dslash$,
which is more difficult to invert than $\Dslash$ itself, as has been shown many
times in the calculation of the quark propagator (BiCGStab is more 
efficient than CG; see, e.g., \cite{BiCGStab}).

\subsubsection{Non-hermitian Chebyshev polynomial}
\label{non-herm}

A rigorous bound of the type Eq.(\ref{eq1}) can also be obtained over an elliptic
domain in the complex plane. In particular, one can try to enclose the 
spectrum $\{\lambda\}$ of the Wilson--Dirac operator $({\bf 1} - \kappa M)$ 
inside an ellipse centered at $(1,0)$. If the major and minor semi-axes
are $a < 1$ and $b$, then the relative error $|\lambda P_n(\lambda) - 1|$ 
is bounded by
\be
2 \left(\frac{1 + b}{1 + \sqrt{1 - a^2 + b^2}} \right)^{n+1} \quad.
\ee
A simple ansatz is to take $a=b$. Then the zeroes of $P_n$ are equally
spaced around the circle of radius $1$ centered at $(1,0)$. 
An example is shown in Fig. 1a, where the boundary of the Dirac spectrum
was estimated from the eigenvalues of the tridiagonal matrix obtained
via BiCGStab.

This approximation converges better than the hermitian one of Sec. \ref{herm} 
\cite{nonherm}. Its main drawbacks are: $(i)$ it is rather unusual (but not
difficult) to monitor the whole complex boundary of the Dirac spectrum
instead of just its smallest eigenvalue; 
$(ii)$ as in Sec. \ref{herm}, the same weight
is given to errors anywhere in the spectrum, whereas the density of 
eigenvalues and the impact of an error vary greatly over
the spectrum. 

\subsubsection{Least-squares polynomial}

Instead of minimizing the maximum relative error, i.e., its infinite norm,
another norm can be chosen. The Euclidean norm $|| R ||_2$
has the advantage of leading 
to a quadratic minimization problem. Once the approximation interval
(e.g., $[\varepsilon,1]$) has been selected, the coefficients of polynomial
$P_n$ can be obtained straightforwardly by solving a least-squares problem
\cite{Montvay}.
There is little additional complication if the domain of approximation is
a rectangle in the complex plane: $[x_1,x_2] \times [y_1,y_2]$.
Besides approximations of $1/x$, other polynomial approximations like
$1/\sqrt{x}$ can easily be constructed. In principle, the weight function 
for the error can be varied as desired over the approximation domain.
Therefore, this scheme offers good flexibility. However, it highlights the 
arbitrariness of what one really tries to optimize.

\subsubsection{Adapted polynomial}

This issue of arbitrariness is addressed in \cite{UV}. 
What one wants is to make the average
correction factor $\Langle \det^2~(\Dslash + m)~P_n(\Dslash + m) \Rangle$
as close to $1$ as possible. This correction factor, or rather its inverse,
can be expressed as a Gaussian integral
\be
\det^{-2}~W = \frac{\int d\eta^\dagger d\eta e^{-\eta^\dagger W^\dagger W 
\eta}}
{\int d\eta^\dagger d\eta e^{-\eta^\dagger \eta}}
= \Langle e^{- |W\eta|^2 + |\eta|^2} \Rangle_{\eta} \quad,
\label{correction}
\ee
where $W = (\Dslash + m)~P_n(\Dslash + m)$ and $\eta$ is a Gaussian random vector.
Since the correction factor must be close to $1$, the Taylor expansion of
the exponential above may be truncated to its first term, yielding
\be
\det^2~W - 1 \approx \Langle |W\eta|^2 - |\eta|^2 \Rangle_{\eta}
\ee
to be minimized.
In \cite{UV}, this requirement is replaced by the sufficient condition
$\parallel W\eta - \eta \parallel^2 \approx 0~\forall~\eta$, 
and $P_n$ is found as the polynomial which minimizes  
$\langle \parallel W\eta - \eta \parallel^2 \rangle_{\eta}$,
by quadratic minimization,
for a fixed set of Gaussian vectors $\eta$. 
The resulting polynomial depends little on the sample of Gaussian vectors,
so that it appears to be adequate to consider only one vector $\eta$.
Moreover, \cite{UV} considers only one representative
background gauge configuration on which to perform the minimization.
In principle, some subtle averaging over gauge configurations should be performed;
in practice, different equilibrium gauge fields yield almost identical
polynomials. This lack of averaging seems to be the only drawback of this
approach, which otherwise takes into account the complete spectral properties
of $\Dslash$, not just its extreme eigenvalues or its spectral boundary.
An illustrative example is shown in Fig. 1b. With $n=16$ the error is as small
as in the approach of Sec. \ref{non-herm} with $n=20$.

All four possibilities reviewed here for choosing $P_n$ converge
exponentially: the relative error decreases as $e^{-c\,n}$, where $c$ depends
on the choice Sec. 2.1.1--4, but goes to zero as $m$, the quark mass.
This implies that, for a constant accuracy, the number of fields $n$ should
grow as $1/m$. 

\subsection{How to make the algorithm exact?}
\label{sect_exact}

Sampling with the measure $\propto |\det^{-1}~P_n(\Dslash + m)|^2$
is an approximation, which deviates from the exact measure 
by a factor $C \equiv |\det~(\Dslash + m)~P_n(\Dslash + m)|^2$.
Several proposals have been made to make the sampling
exact.

\subsubsection{Correction factor}

L\"uscher \cite{Bible} originally suggested that the factor $C$
could be incorporated as a correction in the observable: 
\be
\Langle {\mathcal O} \Rangle_{exact} = \frac{\Langle {\mathcal O}~C \Rangle}
{\Langle C \Rangle} \quad,
\ee
where the right-hand side involves averages over the approximate distribution.
The advantage is that $C$ only needs to be computed when a measurement is
taken, not at every MC step. Sampling can be considered sub-optimal,
since it is performed with respect to the approximate measure; 
however, this can sometimes be turned into an advantage, as for instance
when over-sampling of Dirac near-zero modes is desired.
In any case, the main problem is to compute $C$.

\subsubsection{Lanczos and Metropolis}
\label{nbfields2}
\label{trajectory}

This problem was tackled in \cite{R77,R75}. Since 
$\sqrt{C} = \prod_i~\lambda_i~P_n(\lambda_i)$,
this correction factor can be calculated through a Lanczos process which 
produces all eigenvalues $\lambda_i$ of $(\Dslash + m)$. The error is typically
dominated by the outer (especially the smaller) eigenvalues, which are
determined first, and so the Lanczos process can be stopped early. 
The resulting correction $C$ was used in \cite{R77,R75} to perform an 
accept/reject step and restore importance sampling with respect to the 
exact measure as follows: 
\begin{enumerate}
\item{} Starting from $\{U,\phi\}$, perform a {\em reversible} sequence
of MC steps to obtain $\{U',\phi'\}$;
this sequence is called a trajectory, by analogy with HMC.
\item{} Accept $\{U',\phi'\}$ with probability $min(1,C'/C)$; otherwise,
include $\{U,\phi\}$ in the Markov sequence again.
\end{enumerate}

Reversibility in step (1) means that the probability of choosing the reverse
sequence of local MC updates is equal to that of choosing the original sequence.
This condition is sufficient to ensure detailed balance of the move
$\{U,\phi\} \rightarrow \{U',\phi'\}$ with respect to the approximate MB action
Eq.(\ref{MBaction}) \cite{R57}.

As the lattice volume $V$ is increased, the number of eigenvalues
contributing to $C$ increases like $V$. To keep the acceptance constant,
the relative accuracy of the polynomial approximation must increase
accordingly. Since the approximation converges exponentially in $n$, 
this implies that $n \propto \log V$.

\subsubsection{Stochastic Metropolis}

In the acceptance test above, $C'/C$ is only compared with a random number.
Therefore, it should not be necessary to compute this factor exactly: an estimate
should suffice. As in Eq.(\ref{correction}), one may readily see that
$\frac{C'}{C} = \Langle e^{- |W'^{-1} W \eta|^2 + |\eta|^2} \Rangle_\eta$,
where $W = (\Dslash + m)~P_n(\Dslash + m)$ and $\eta$ is a Gaussian random
vector. Therefore, step $(2)$ above can be replaced by a noisy Metropolis
test \cite{KenKuti} as:
\begin{enumerate}
\setcounter{enumi}{1}
\item Draw a Gaussian random vector $\eta$, and 
accept $\{U',\phi'\}$ with probability 
$min(1, e^{- |W'^{-1} W \eta|^2 + |\eta|^2})$.
\end{enumerate}
This is a very economical way to enforce exactness of the algorithm.
Since $W'$ is close to unity, $W'^{-1} (W \eta)$ can be computed in a few
iterations of a linear solver. It is even possible to cut down on the 
number of iterations by first drawing the random number in the acceptance test
\cite{R20}.
In case $W$ is difficult to express, as for instance for $1$ flavor,
where $W = (\Dslash + m)^{1/2}~P_n(\Dslash + m)$, 
it can be replaced by a high-quality polynomial approximation
$W \approx P_{\tilde{n}}^{-1}(\Dslash + m)~P_n(\Dslash + m)$,
where ${\tilde{n}} \gg n$ \cite{Nf1}.
A similar strategy is used in \cite{SUSY}.

The average acceptance depends on the fluctuations of $C'/C$ about 1.
It can be predicted rather accurately as a function of $n$, $m$, and $V$,
with a single-parameter ansatz accounting for the fluctuations of the
gauge field (i.e., for the value of $\beta$) \cite{R57}. As $n$ varies, the
acceptance behaves as $e^{-e^{-n}}$, so that it changes rather abruptly
from nearly zero to nearly 1.

\subsection{Reducing the degree of the polynomial by preconditioning}

The difficulty of approximating $1/x$ increases with the width of the
approximation interval. As can be seen already from Eq.(\ref{eq1}),
the degree $n$ of the approximation must increase linearly with the
``dynamic range'' $\lambda_{max} / \lambda_{min}$ $(= 1 / \varepsilon$
for Sec. \ref{herm}) considered.
Reduction of this dynamic range, or equivalently, of the condition number of the
matrix to invert, may be achieved by preconditioning.
Two simple and efficient ideas have been put forward.

\subsubsection{Even--odd preconditioning}

This preconditioning exploits the identity $\{M,\Sigma\}=0$, where $M$
is the Wilson (or staggered) hopping matrix, and $\Sigma$ is the diagonal
matrix with elements $\Sigma_{x,y} = (-1)^x \delta_{x,y}$. It follows that
\be
\det({\bf 1} - \kappa M) = \det({\bf 1} + \kappa M) 
= \det({\bf 1} - \kappa^2 M_{eo}M_{oe}) \quad,
\ee
where $M_{eo}$ hops from odd sites to even ones. The preconditioned matrix
$\hat{\Dslash} \equiv {\bf 1} - \kappa^2 M_{eo}M_{oe}$
is easier to invert than the original one, as is well known in the context
of quark propagator calculations. Now the approximating polynomial
$P_n(\hat{\Dslash})$ factorizes into 
$\prod_k (\hat{\Dslash} - \hat{z_k} {\bf 1})$,
and each monomial gives rise to a partial determinant 
$\det^{-1} (\hat{\Dslash} - \hat{z_k} {\bf 1})^\dagger 
(\hat{\Dslash} - \hat{z_k} {\bf 1})$.
Turning this determinant into a Gaussian integral, one would expect a
multiboson action 
$|(\hat{\Dslash} - \hat{z_k} {\bf 1}) \phi_k|^2$, i.e., with range $4$.
Fortunately, a simpler Gaussian integral can be used, because
\be
\det({\bf 1} - \kappa^2 M_{eo}M_{oe} - \hat{z_k} {\bf 1}) =
\det\left(\begin{array}{cc}
{\bf 1 - x} & ~~- \kappa M_{eo} \\
- \kappa M_{oe} & ~~{\bf 1 - y}
\end{array}\right) \quad,
\ee
provided 
$(1-x)(1-y) = 1 - \hat{z_k}$.
Thus, for the effective action one may use
$|({\bf 1} - z_k - \kappa M) \phi_k|^2$,
where $z_k$ takes value $x$ on even sites and $y$ on odd ones, satisfying
the above equality.
The simplest choice is 
$x=y=1 - (1 - \hat{z_k})^{1/2}$ \cite{R57}; alternatively,
$y=0, x=\hat{z_k}$ has also been used \cite{R60}.
Either way, the degree $n$ of the approximation is reduced by a factor of 2
or more, with no overhead.

\subsubsection{UV filtering}
\label{UV-filtering}

This preconditioning makes use of the identity
$e^{-{\rm Tr}A}~\det~e^A = 1$,
so that 
\be
\det({\bf 1} - \kappa M) \equiv 
e^{- \sum_j a_j {\rm Tr} M^j} \times
\det\left( ({\bf 1} - \kappa M) ~~e^{+ \sum_j a_j M^j} \right) \quad.
\ee
The number of non-zero coefficients $a_j$ and their values can be adjusted
and optimized freely. The idea is that the large Dirac eigenvalues, 
corresponding to UV Fourier modes, can be accounted for by the first factor
$e^{- \sum_j a_j {\rm Tr} M^j}$.
The polynomial $P_n(M)$ must then approximate 
$({\bf 1} - \kappa M)^{-1} ~~e^{- \sum_j a_j M^j}$,
where the UV modes have been exponentially damped, or ``filtered out.''
The dynamic range of the approximation is thus reduced, and with it the
degree of the polynomial approximation.
There is no overhead for this preconditioning up to order $M^4$: 
the first three terms give zero trace, and the fourth simply shifts the
coupling $\beta$ of the plaquette in the gauge action.
The values of the coefficients $a_j$ can be optimized at the same time
as the polynomial $P_n$ is determined (see \cite{UV} for details).
This UV filtering amounts to removing from the determinant the first term(s)
of its loop expansion. It has been observed that the major effect of
dynamical quarks is to shift the gauge coupling $\beta$, and not much else,
down to rather light quarks \cite{AnnaH}. Therefore, the preconditioning
should be very effective. Indeed, the number of boson fields remains 
much smaller than the number of linear solver iterations necessary for HMC
with the same parameters, and the memory bottleneck of the MB approach
essentially disappears. Figure 1c shows the zeroes of a degree-7 polynomial which
provides the same accuracy as those of Figs. 1a and 1b. Note how these
zeroes are concentrated near the IR part of the spectrum.

\begin{figure*}[!thb]
\vspace{0.5cm}
\begin{center}
\epsfig{file=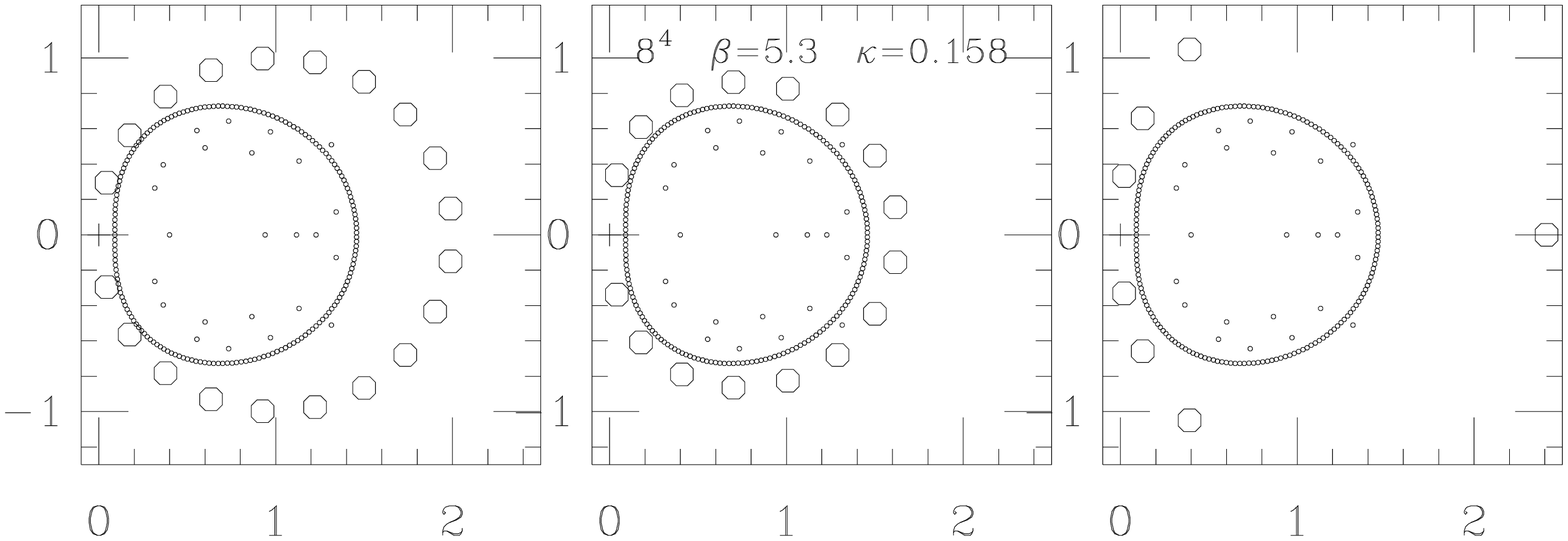,height=5cm,width=14cm,angle=0}
\caption{Three strategies to choose the zeroes of the
approximating polynomial, all giving similar acceptances: 
from left to right, (a) non-hermitian Chebyshev 
polynomial ($n=20$); (b) adapted polynomial ($n=16$); (c) adapted polynomial
with UV-filtering ($n=7$). The small symbols, forming a contour, are 
an estimate of the boundary
of the complex Dirac spectrum, obtained as a by-product from the iterative 
solver.}
\end{center}
\end{figure*}

\subsection{Implementation details}

As seen in Sec. \ref{sect_exact}, 
making the MB method exact involves applying the polynomial
$P_n(\Dslash)$ to some vector $\vec{v}$. Doing this accurately enough can
be problematic on 32-bit machines like the Quadrics/APE, and
much effort has been devoted to the study of round-off errors 
\cite{Montvay,ordering}.
The lessons learned from these studies are: 
$(i)$ whenever possible, avoid representing
$P_n$ as a product of monomials. $P_n$ can be decomposed on a
basis of orthogonal (esp. Chebyshev) polynomials, where at each stage
the lower-degree polynomial is the best approximation to $P_n$. 
The so-called Clenshaw recursion \cite{NumRec,ordering} gives particularly
good control over round-off errors; $(ii)$ on 32-bit machines, applying
$P_n$ naively as a product of monomials will give unacceptable errors even
for moderate $n$ (20--30), unless careful ordering of the factors is performed.
Montvay's ordering \cite{Montvay} and the bit-reversal scheme \cite{ordering}
appear to be best.

\section{Dynamics}

\subsection{Theory}
\label{autocorr}

The theoretical understanding of MB dynamics is very crude.
The situation is more complicated than with HMC, because one normally
alternates single MC steps for the gauge fields and for the boson fields,
so that the two dynamics are intimately coupled. Nevertheless, some simple
facts derive from the form Eq.(\ref{MBaction}) of the MB effective action.
\begin{enumerate}
\item Each boson field contributes a harmonic piece to the potential seen
by the gauge field, and this potential gets steeper with the number $n$
of boson fields. Consequently, the step size of the gauge-field update
decreases as $1/n$, and the autocorrelation time increases as $n$ \cite{R77,R78}.
\item When the gauge field is kept fixed, each boson species $\phi_k$ has a 
mass which is the smallest singular value of $({\bf 1} - z_k - \kappa M)$.
This mass goes to zero as $z_k$ approaches the boundary of the Dirac spectrum.
This normally happens principally to the boson fields governing the IR part
of the spectrum, which therefore have the slowest dynamics.
\end{enumerate}
It is thus clear that critical slowing down occurs in both the gauge and 
the boson sectors as the quark mass is decreased. Under plausible assumptions,
Ref. \cite{R57} suggests an autocorrelation time $\sim n / m^z$, where
$z = 1-2$ is the dynamical critical exponent for the boson-field update.

\subsection{Practical knowledge}

Because the MB action is local, a choice of standard local MC updates is
available: Metropolis, heatbath, and overrelaxation, for both gauge and
boson fields [pseudo-heatbath for $SU(3)$ gauge fields].
A trajectory between successive global accept/reject steps 
must consist of a reversible sequence of local MC steps (Sec. \ref{trajectory})
Beyond this, there is complete freedom in choosing among 
the various updates.

It has been observed that a judicious ratio of gauge to boson
updates and of heatbath to overrelaxation can
accelerate the dynamics by a factor of $\sim 3$ \cite{R60}. However, this tuning
must presumably be performed anew for each value of the quark mass 
(and preferably for each boson species): no general rule has emerged.

Also, since the frozen bosons prevent a large MC step for the gauge fields,
a combined update (one link + all boson fields at its two ends) has been
devised \cite{R78,R60}. However, the gain in the dynamics is moderate,
and is all but absorbed in the extra work per update.

\subsection{Implementation}

Since the MB action is local, it is trivial to divide the lattice into sets
of decoupled variables, and to distribute the update of a set on a parallel
machine. MB programs have been written for the Quadrics/APE, the T3D/E,
or based on MPI for portability. The minimum decomposition for Wilson
fermions ($r_{{\rm Wilson}} = 1$) is into 8 sets (2 opposite points per $2^4$ 
hypercube \cite{R57}). A more efficient organization is the ``star''
updating strategy \cite{star}: all boson fields at a given site and all
8 gauge links attached to it are updated before going to the next site.
This allows reuse of intermediate results (e.g., the corners of the gauge
plaquettes), thus reducing the total amount of work. And one can easily
integrate analytically over the boson fields at the central site, which 
yields an effective gauge action allowing for a larger MC step.
This provides advantages similar to those of the combined update of \cite{R78}
above,
without the overhead.

\subsection{Prospects}

Overrelaxation \cite{Adler} is a remarkably efficient local update.
There is little hope of improving on it. On the other hand, the action of
each boson field is Gaussian. It is so simple that one may hope to 
accelerate the boson dynamics at low computing cost. One avenue still not
explored is to devise a cluster MC. A simpler one is to consider a global
boson heatbath, achieved by assigning 
$\phi_k = ({\bf 1} - z_k - \kappa M)^{-1} \eta$,
with $\eta$ a Gaussian random vector. Theoretical arguments have been proposed,
suggesting that such a global update is superior to a local one as 
$m \rightarrow 0$ \cite{R55}. The recent proposal of a quasi-heatbath 
\cite{QHB}
(approximate global heatbath + accept/reject) reduces the cost of such
a strategy and makes it worth exploring.

\section{Assessment of MB vs HMC}

\subsection{Theoretical cost analysis}

A naive analysis of the cost of a MultiBoson simulation as a function
of the volume $V$ and the quark mass $m$ goes as follows \cite{R57,R55,Bielefeld}.
\begin{itemize}
\item The work per bosonic update step is proportional to $V$ and to the 
number $n$ of boson fields. The same holds for a gauge update step, because
the gauge force (``staple'') sums up $n$ boson contributions.
\item The autocorrelation time grows as $n / m^z$, with $z=1$---2 
(see Sec. \ref{autocorr}).
\item The number of fields $n$ grows as $m^{-1} \log V$ 
(see Secs. \ref{nbfields1} and \ref{nbfields2}).
\end{itemize}
The work per independent configuration thus grows as $V (\log V)^2 m^{-2-z}$,
to be compared with the HMC case: $V^{5/4} m^{-p}$, $p = 13/4$---$7/2$ 
\cite{ADK,Bielefeld}.
It is likely that the boson dynamics have a dynamical exponent $z=2$.
In that case, the MB approach would scale somewhat better than HMC with respect 
to the volume, and rather worse with respect to the quark mass.
This analysis is no more than plausible. But it agrees with practical 
observations: with MB simulations, one tends to pay a low price when 
increasing the lattice size, and a rather stiff one when decreasing the
quark mass.

\subsection{Pros and cons of MB}

Some of the more or less well known virtues and shortcomings of the
MB approach are the following. 

Pros:
\begin{itemize}
\item As the fermion mass $m$ increases, the number $n$ of boson fields can
be reduced, and the MB dynamics approach quenched
local dynamics (pseudo-heatbath and overrelaxation). 
This is in contrast to HMC, which approach quenched molecular dynamics,
measured to be almost two orders of magnitude slower \cite{MDdynamics}.
Therefore, MB has to be much faster than HMC if the fermions are heavy enough.
\item The MB action is quadratic in the fermion mass $m$ or in $\kappa$.
This allows for straightforward reweighting of the Monte Carlo results for
[slightly] different masses.
\item The enlarged MultiBoson phase space facilitates the dynamics around
effective energy barriers. 
\end{itemize}
This last notion must be clarified. Consider the gauge force 
$\frac{\partial S}{\partial U}$ in the MB action 
$S = \sum_k |(\Dslash(U) + m - z_k) \phi_k|^2$.
Each of the $n$ terms in the derivative contains a {\em different} factor of
$\phi_k^\dagger \phi_k$. These factors add up incoherently, in contrast to
HMC where there is only {\em one} factor of $\phi^\dagger \phi$. This 
incoherent sum has the effect of smoothing the divergence of the gauge force
in the vicinity of a Dirac zero mode. 
Equivalently, the added bosonic dimensions in the phase space provide the
MB algorithm with many paths which ``go around'' the energy barrier,
making its crossing easier.
This may give a qualitative advantage
to the MB method for moving through topological sectors at small quark masses.
Similarly it is the {\em decoupling} of the various Fourier modes which makes 
UV filtering possible (Sec. \ref{UV-filtering}). If one uses the same UV-filtered
polynomial in the HMC action, the step size must be adjusted to track
the fast dynamics of the UV modes, which makes the IR dynamics hopelessly slow.

Cons:
\begin{itemize}
\item MB needs more memory to store the bosonic fields. However, this no longer
represents a real obstacle.  With UV-filtering, simulations
of large volumes at small masses can be performed with no more than 
${\mathcal O}(50)$ boson fields.
\item The MB approach is designed for the standard (Wilson or staggered)
Dirac discretization. Any less local discretization makes it very difficult
to use efficient local update algorithms for the gauge and boson fields.
\end{itemize}

\subsection{Actual performance comparison}

Performance comparisons with HMC are summarized in chronological order in the
rightmost column of Table I.
There is a slight trend for the advantage of MB
to increase with time owing to successive algorithmic improvements.
But several caveats are in order.
\begin{enumerate}
\item Extracting autocorrelation times with some accuracy is notoriously
expensive. Moreover, with MB simulations, there is a systematic effect by
which unusual configurations, in the tail of the distribution, tend to
be associated with much longer autocorrelation times because the polynomial
$P_n$ is a bad approximation on such background fields and the acceptance
(Sec. \ref{trajectory}) is very small.
Therefore, short MC runs tend to paint a rosy picture by underestimating
the MB autocorrelation time.
\item The version of HMC serving for comparison does not always incorporate
the complete set of ``bells and whistles'' (even--odd or SSOR preconditioning,
BiCG solver, incomplete convergence or extrapolation of solution during
MD trajectory \cite{ADK}). Each of these technical refinements improves
HMC performance by a factor ${\mathcal O}(2)$.
\item As seen above, MB is a superior method when the fermions are heavy.
So a meaningful question would be: how small a fermion mass does it take
for HMC to win over MB, if at all? But, of course, light fermion simulations
are even more expensive.
\item What is compared is usually the integrated autocorrelation time of
the plaquette, expressed in applications of the Dirac matrix to a vector.
Sometimes, more relevant, larger-scale observables like the pion correlator
have also been measured, and give rather similar results. 
However, a comparison of the decorrelation of the topological charge,
yet to be performed, might present a different picture.
\end{enumerate}

To summarize, my subjective assessment of Table I is that MB is finally
becoming competitive with HMC in the interesting regime of quark masses.

\section{Conclusion}
 
The initial enthusiasm for the MB method has now abated. The general impression
is that for light quarks, MB is roughly equivalent to HMC in terms of efficiency, and that
research effort is better spent on improving the discretization of the Dirac
operator. Indeed, one might argue that everything that can be done with MB
can be done, perhaps better and more simply, within the HMC framework.
 
MB replaces the {\em adaptive} inversion of the Dirac operator performed by
the linear solver at each HMC step by a {\em fixed} approximation. This
substitution can be performed directly in the HMC effective action
\cite{Taka}. While it leads to no special advantage for the simulation of
2 flavors, this replacement allows the {\em exact} simulation of any number
of flavors ,
(with the same limitations as MB on the positivity of the Dirac eigenvalues
\cite{Nf1}),
and opens the possibility of biased sampling, e.g., of Dirac near-zero modes
\cite{PHMC}.
 
However, this dismissive statement is not completely true. The presence of multiple
boson fields changes the dynamics significantly. It essentially replaces
the {\em deterministic}, IR-driven dynamics of HMC by fast {\em diffusive}
dynamics. This is a clear advantage for heavy fermions, although possibly
a handicap for light ones. In addition, the multiple boson fields allow 
the breaking and separate treatment of the various scales (IR and UV) of the Dirac
operator, as in the UV-filtered version. A similar treatment is hopelessly
inefficient within the HMC framework.
 
Nevertheless, the strongest limitation of the MB approach appears to be its
most highly praised characteristic: the action is {\em local}. Actually,
the range of the boson interaction is twice that of the Dirac operator.
This makes it cumbersome to implement local update steps for any but the
simplest Dirac discretization (Wilson or staggered). As increasing attention
is paid to reducing discretization errors, the Dirac operator becomes less
local and the MB approach loses much of its appeal.
 
Therefore, the MB method at present is a ``niche'' algorithm.
It is a superior choice for the simulation of heavy dynamical fermions.
And the same algorithm ``template'' can be used to simulate exotic numbers
of flavors or other unusual fermionic determinants. Further research
efforts to accelerate MB dynamics may well broaden its appeal.

\bibliography{}

\end{document}